\documentclass[12pt]{spieman}  
\usepackage{amsmath,amsfonts,amssymb}
\usepackage{graphicx}
\usepackage{setspace}
\usepackage{tocloft}
\usepackage{gensymb}
\usepackage{enumitem}
\usepackage{xcolor}
\setlist{  
  listparindent=\parindent,
  parsep=0pt,
}
\usepackage{rotating}
\usepackage{float}
\usepackage{multirow}
\usepackage{lineno}
\usepackage{adjustbox}

\title{Implementing multi-wavelength fringe tracking for the Large Binocular Telescope Interferometer's phase sensor, PHASECam}

\author[a,*]{Erin R. Maier}
\author[b]{Phil Hinz}
\author[c]{Denis Defr\`{e}re}
\author[a]{Paul Grenz}
\author[a]{Elwood Downey}
\author[d,a]{Steve Ertel}
\author[a]{Katie Morzinski}
\author[a]{Ewan S. Douglas}
\affil[a]{Steward Observatory, Department of Astronomy, University of Arizona, 933 N. Cherry Ave, Tucson, AZ 85721, USA}
\affil[b]{Department of Astronomy, University of California Santa Cruz, Laboratory for Adaptive Optics, Santa Cruz, CA 95064, USA}
\affil[c]{STAR Institute, Universit\'{e} de Li\`{e}ge, 17 Alle\'{e} du Six Ao\^{u}t, B-4000 Sart Tilman, Belgium}
\affil[d]{Large Binocular Telescope Observatory, 933 North Cherry Avenue, Tucson, AZ 85721, USA}

\cftpagenumbersoff{figure}
\cftpagenumbersoff{table} 
\begin{document} 
\maketitle

\begin{abstract}
PHASECam is the fringe tracker for the Large Binocular Telescope Interferometer (LBTI). It is a near-infrared camera which is used to measure both tip/tilt and fringe phase variations between the two adaptive optics (AO) corrected apertures of the Large Binocular Telescope (LBT). Tip/tilt and phase sensing are currently performed in the $H$ (1.65 $\mu$m) and $K$ (2.2 $\mu$m) bands at 1 kHz, but only the $K$-band phase telemetry is used to send corrections to the system in order to maintain fringe coherence and visibility. However, due to the cyclic nature of the fringe phase, only the phase, modulo 360$\degree$, can be measured. PHASECam's phase unwrapping algorithm, which attempts to mitigate this issue, occasionally fails in the case of fast, large phase variations or low signal-to-noise ratio. This can cause a fringe jump, in which case the OPD correction will be incorrect by a wavelength. This can currently be manually corrected by the operator. However, as the LBTI commissions further modes which require robust, active phase control and for which fringe jumps are harder to detect, including multi-axial (Fizeau) interferometry and dual-aperture non-redundant aperture masking interferometry, a more reliable and automated solution is desired. We present a multi-wavelength method of fringe jump capture and correction which involves direct comparison between the $K$-band and $H$-band phase telemetry. We demonstrate the method utilizing archival PHASECam telemetry, showing it provides a robust, reliable way of detecting fringe jumps which can potentially recover a significant fraction of the data lost to them.
\end{abstract}


{\noindent \footnotesize\textbf{*}Erin Maier, \linkable{erinrmaier@email.arizona.edu}} 

\begin{spacing}{1}   

\section{Introduction}
\label{sect:intro}  

A significant obstacle faced by ground-based optical and infrared interferometry is rapid Optical Path Difference (OPD) variations between telescope apertures introduced by turbulence in the atmosphere as well as mechanical sources such as telescope vibration. This leads to a temporal loss of coherence between the wavefronts and prevents meaningful measurements of the contrast of any given fringe, leading to reduced accuracy and precision of visibility measurements. 

One way to mitigate the loss of coherence is to use a fringe tracker\cite{Shao80,Colavita99}. Fringe trackers are devices which measure and correct the OPD in real time. Some keep the fringe packet approximately centered to within a fraction of the coherence length, usually a few wavelength: these are considered ``coherencers". Others such as the GRAVITY instrument at the Very Large Telescope Interferometer (VLTI) and the Keck Interferometer track the fringe phase delay in order to reduce the OPD to a fraction of the observing wavelength\cite{Lacour19,Colavita10}. 

However, when phase tracking is performed using quasi-monochromatic fringes, there exists a 360$\degree$ (2$\pi$) degeneracy in the phase measurement due to the cyclic nature of the fringe phase. If a fringe measurement is bad, due to effects such as fast OPD variations, low SNR due to extreme tip/tilt, etc., phase variations on the scale of $\lambda$/2 may be missed - or wrongly detected - due to the degeneracy. This leads to fringe jumps, i.e., unsensed shifts into adjacent fringes. This can further lead to significant loss of achievable fringe contrast due to the combination of different contrasts across an observation or even an individual integration. The development of methods to detect and correct fringe jumps is thus of ongoing interest to developers of interferometric instruments. A class of methods of particular use involves multi-wavelength phase sensing. As fringe phase is wavelength dependent, measurements at multiple wavelengths will break the phase ambiguity. Multiple techniques can be used simultaneously: various modern interferometers\cite{Colavita10,Lacour19} combine group delay with phase delay tracking, and coherence envelope tracking can also be used with phase delay tracking. Multi-wavelength methods are also used in applications such as optical metrology to create a larger synthetic wavelength\cite{Wyant71,Polhemus73}.


\begin{figure}[ht!]
\centering
\includegraphics[width=\textwidth]{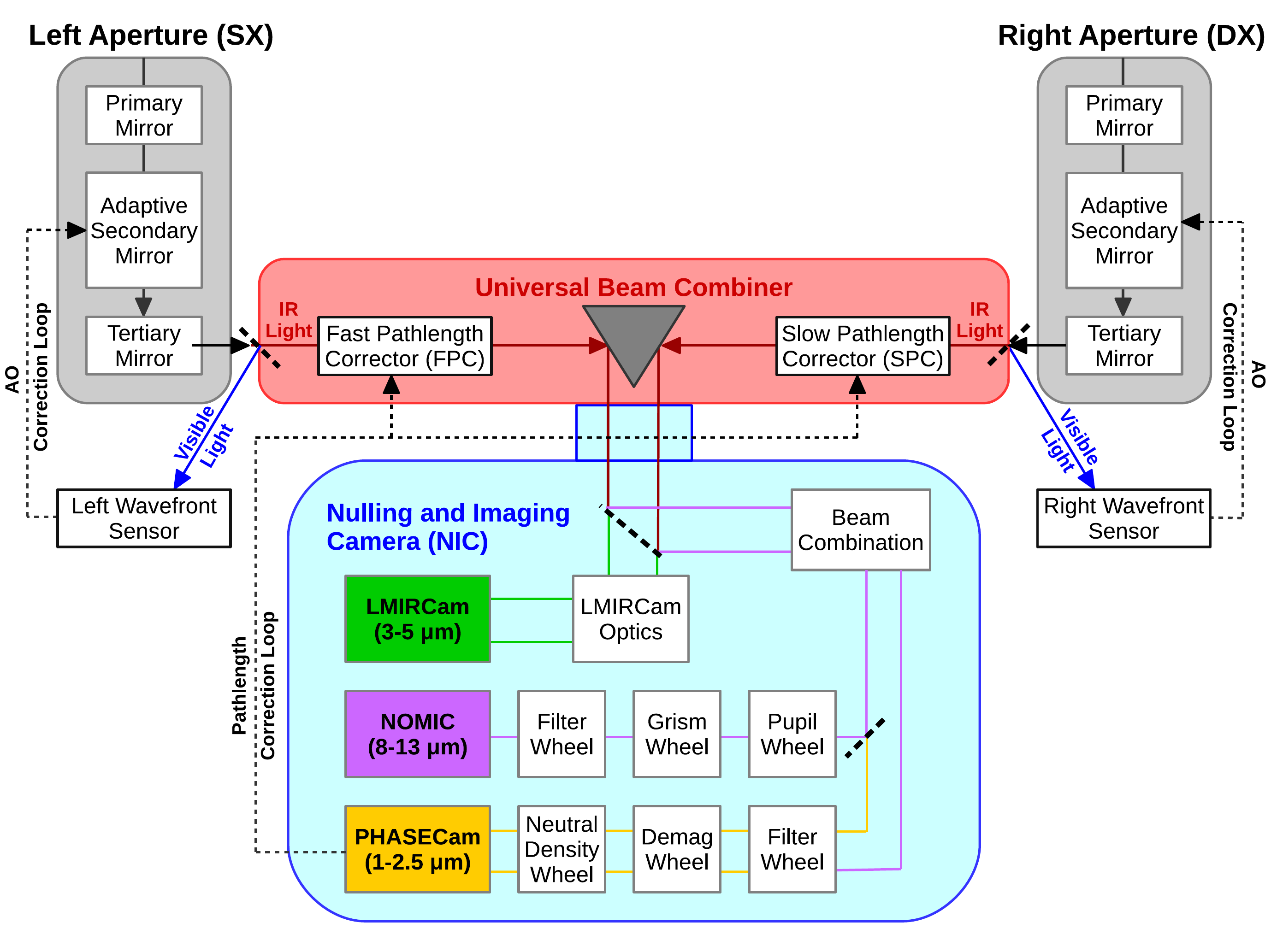}
\caption{System-level block diagram of the Large Binocular Telescope Interferometer (LBTI) showing the optical light path through the telescope and the instrument, including the Universal Beam Combiner (UBC, red) and the Nulling and Imaging Camera cryostat (NIC, blue). Visible light is reflected from the UBC entrance window to be used for wavefront sensing by the telescope, while infrared light passes into the cryogenic LBTI system. The beam combiner directs the light into the NIC cryostat, where the thermal infrared light (green, 3-5 $\mu$m) is directed to LMIRCam (the $L$ and $M$ InfraRed Camera) for imaging and Fizeau interferometry, the mid-infrared (purple, 8-14 $\mu$m) light is directed to NOMIC (Nulling Optimized Mid-Infrared Camera) for nulling interferometry, and the near-infrared (orange, 1.5-2.5 $\mu$m) light is directed to PHASECam for tip/tilt and phase sensing. PHASECam receives both outputs of the beam combiner, and sends tip/tilt and OPD corrections to the Fast and Slow pathlength correctors in the UBC (FPC/SPC). This diagram is schematic only and does not show every optic. Adapted from Defr\`{e}re et al (2016a).}
\label{system-block}
\end{figure}

The Large Binocular Telescope Interferometer (LBTI)\cite{Hinz16, Hinz18} is a NASA-funded nulling and imaging instrument designed to coherently combine the two primary mirrors of the Large Binocular Telescope (LBT)\cite{Hill14, Veillet14, Veillet16} for high-sensitivity, high-contrast, and high-angular resolution infrared (1.5 - 13 $\mu$m) imaging and interferometry. The LBTI is equipped with two science cameras: LMIRCam\cite{Wilson08, Skrutskie10, Leisenring12} (the $L$ and $M$ InfraRed Camera, 3-5 $\mu$m) and NOMIC\cite{Hoffmann14} (Nulling Optimized Mid-Infrared Camera, 8-14 $\mu$m). The LBTI's fringe tracker is PHASECam\cite{Defrere14}, a near-infrared (1.5-2.5 $\mu$m) camera which measures and corrects differential OPD and tip/tilt variations between the two adaptive optics (AO)-corrected apertures of the LBT. A block optical path diagram of the telescope optics, LBTI's Universal Beam Combiner\cite{Hinz04} (UBC) and the Nulling and Imaging Camera\cite{Hinz08} (NIC) cryostat where PHASECam, NOMIC, and LMIRCam are housed can be seen in Fig. \ref{system-block}\cite{Defrere16a}.

Previously, PHASECam operated as coherencer. It utilized a contrast gradient metric which tracked changes in the fringe contrast in order to determine the OPD and sent corrections at a rate of 1 kHz\cite{Defrere14}. LBTI's first stable on-sky fringes were obtained in December 2013 using this method. However, the contrast gradient is a nonlinear metric, limiting the precision of the correction to a closed-loop residual OPD of approximately 1 $\mu$m\cite{Defrere14}. This precision was not sufficient for the LBTI's primary science cases, the detection and characterization of exozodiacal dust and exoplanets. It also produced a non-Gaussian phase distribution which complicated the use of advanced data reduction techniques\cite{Defrere16a}. Thus, phase delay tracking was required. PHASECam transitioned to phase delay tracking in 2015. It measures the fringe phase in both the $H$ (1.65 $\mu$m) and $K$ (2.2 $\mu$m) bands at 1 kHz. However, it currently only utilizes the $K$-band phase telemetry for active phase control.

\begin{figure}[ht!]
\centering
\includegraphics[width=.8\linewidth]{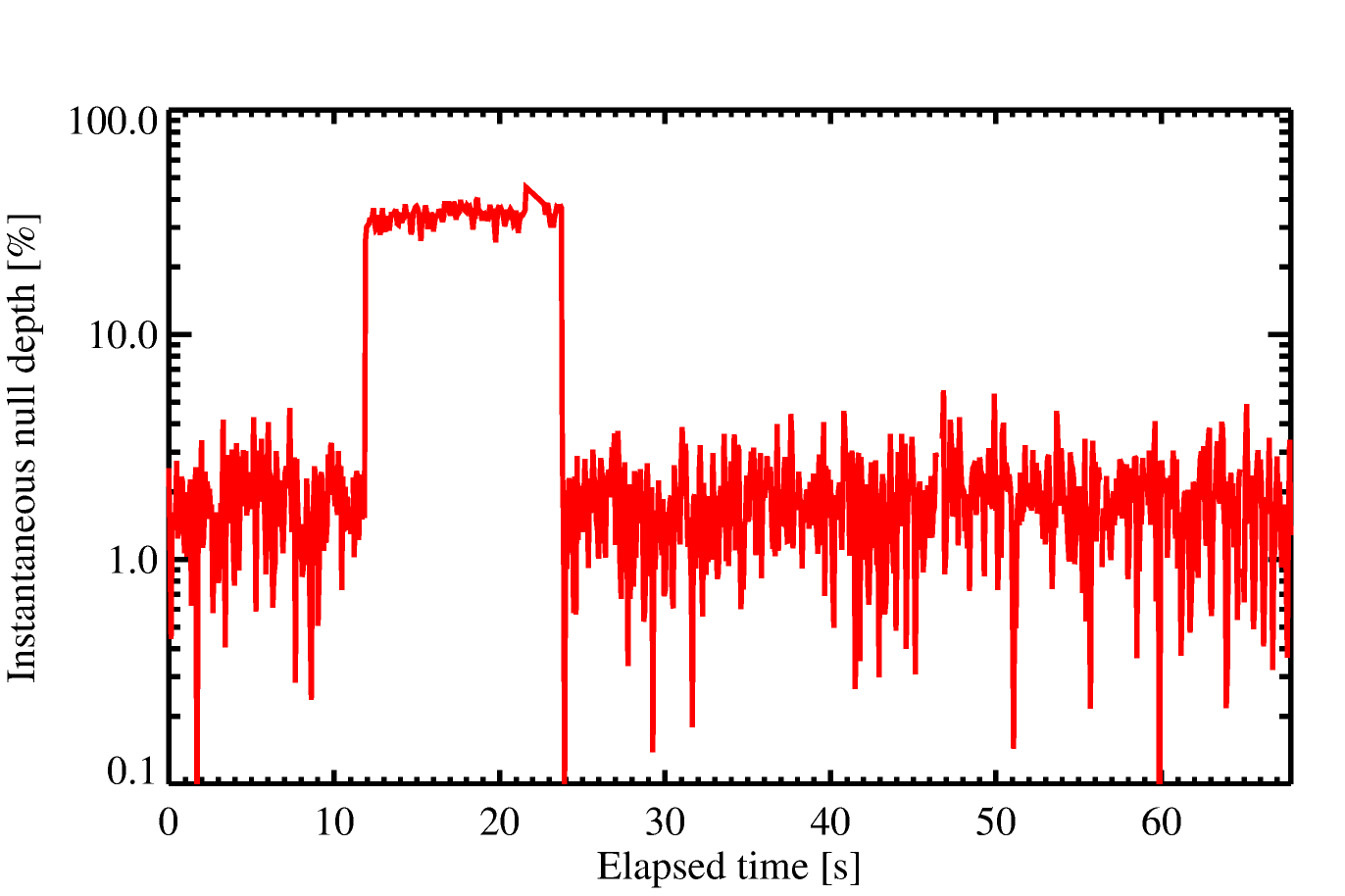}
\vspace{4mm}
\caption{A typical null depth time sequence during nulling observations with NOMIC, showing the sudden jump away from the optimum null depth indicative of a fringe jump at $\sim$0:11.5s.}
\label{null-sequence}
\end{figure}

PHASECam experiences fringe jumps not infrequently, with gaps between successive jumps from on the order of $\sim$100 s to $<$5 s dependent upon a variety of factors. Currently, fringe jumps are corrected manually by the PHASECam operator. The utilization of the $H$-band phase telemetry to implement multi-wavelength fringe tracking has been previously discussed\cite{Defrere14}. However, it has not been a priority as until now PHASECam has primarily been used for the nulling interferometric observations of the Hunt for Observable Signatures of Terrestrial Planets (HOSTS) survey. HOSTS is a NASA-funded $N$-band (10 $\mu$m) survey of exozodiacal dust around nearby stars\cite{Defrere16a, Ertel18a, Ertel18b, Ertel2020}. Fringe jumps during nulling observations are visible in the real-time null telemetry as a departure from the nominal maximum null depth, as can be seen in Fig. \ref{null-sequence}, and can thus be detected and corrected manually in real time. However, in recent observing semesters the LBTI has begun to commission other observing modes which require active phase control, including imaging, or ``Fizeau", interferometry as well as Non-Redundant Aperture Masking interferometry (NRM), thus making automated, reliable fringe jump detection and correction higher priority. 

\subsection{Fizeau interferometry}

Fizeau interferometry at the LBTI utilizes multi-axial image plane beam combination across the entire 22.8-m edge-to-edge LBT mirror separation\cite{Patru17a, Patru17b}. This is in contrast to nulling interferometry, which utilizes co-axial pupil plane beam combination across the 14.4-m center-to-center mirror separation. Previously, LBTI has only imaged a small number of targets in Fizeau mode.\cite{Leisenring14, Conrad15, Conrad16, deKleer17, Hill13} However, these previous observations were in ``lucky" Fizeau mode, whereby the targets were imaged with short integration times and without active phase control and the few ``lucky'' frames where the fringes were well-overlapped and well-centered were selected out during the data reduction phase.\cite{Fried78} ``Lucky" imaging is another method to mitigate the effect of OPD variations, but it can severely limit the sensitivity of observations. There is also no way to know  which fringe has been centered in a given integration, leading to a heavy loss of achievable precision of visibility measurements. 

\begin{figure}[ht!]
\centering\includegraphics[width=.8\textwidth,]{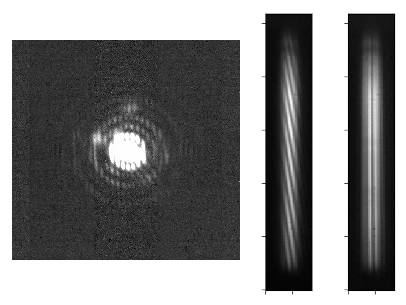}
\caption{ $Left$: The L' dual-aperture Fizeau PSF, with optical ghosts. The Airy ring pattern is that of a single 8.25 m effective diameter aperture, overlaid by fringes corresponding to the 14.4 m center-to-center aperture separation. $Middle$: The spectrally dispersed dual-aperture Fizeau PSF after initial detection of the coherence envelope with LMIRCam. It is dispersed with a 2.8-4.2 $\mu$m $L$-band grism\cite{Kuzmenko12}, creating the ``barber-pole'' Fizeau fringes. The vertical direction is the spectral axis, with bluer wavelengths at the top and redder at the bottom. The slant of the fringes indicates OPD error between the beams which must be removed with PHASECam. $Right$: The approximately vertical Fizeau fringes after minimization of the OPD. Modified from Spalding et al (2018)\cite{Spalding18}.}
\label{phase-lock-modes}
\end{figure}

The primary HOSTS observations concluded during the 2018A observing season\cite{Ertel18b}. The LBTI has since begun to commission phase-controlled Fizeau imaging with LMIRCam.\cite{Spalding18,Spalding19} Active, reliable phase control is critical to maintaining zero OPD alignment of the Fizeau coherence envelope and remaining locked on the central science fringe for the duration of the observation. Fig. \ref{phase-lock-modes} shows a spectrally dispersed Fizeau point spread function (PSF) on the LMIRCam detector immediately after the initial finding of the coherence envelope, as well as after minimization of the OPD using PHASECam. PHASECam has successfully been used to stabilize the science fringes in some Fizeau observations for up to a few minutes since commissioning began. However, the process is still under development, and fringe jumps negatively impact the measurements.

\subsection{Non-redundant Aperture Masking Interferometry (NRM)}

NRM transforms large apertures into multi-element Fizeau interferometers by utilizing a pupil-plane mask to produce non-redundant baseline separations within and between apertures\cite{Leisenring12}. The power measured at certain spatial frequencies and position angles is associated with pairs of mask sub-apertures, allowing for Fourier amplitudes and phases for each of the baselines to be measured.

The addition of single aperture sparse aperture masking capabilities to facilities with AO, such as Keck, Subaru and the Very Large Telescope, has been highly successful.\cite{Tuthill06,Martinache09,Norris15,Cheetham16}. This holds true with the LBT, where NRM observations have produced scientific results with both single and dual apertures without active phase control\cite{Sallum15,Sallum17}. Closed loop phase control is thus a natural step forward which will allow for dual-aperture NRM observations of unprecedented precision, on spatial scales even smaller than the binocular resolution $\left(\frac{\lambda}{22.8 m}\right)$\cite{Hinz16}. The first extended closed-loop dual-aperture NRM observations were obtained using PHASECam in May 2018. As with controlled Fizeau observations, however, these observations suffered from the presence of fringe jumps and would benefit from active detection and correction.

\subsection{Outline}

We have developed an algorithm which implements multi-wavelength fringe tracking for PHASECam, by combining the $K$-band phase telemetry with the $H$-band phase telemetry. The outline of this paper is as follows: we give a more detailed overview of PHASECam and its current approach to phase sensing and fringe jump detection and correction, updated from previous publications\cite{Defrere14, Defrere16a}, in Sec. 2. We describe our multi-wavelength approach, its implementation, and preliminary tests in Sec. 3. We present results of said tests and discussion thereof in Sec. 4, and finally conclude with some discussion of future work and broader applications in Sec. 5.

\section{PHASECam and its Algorithms}

\subsection{Overview}\label{overview}

PHASECam uses a fast-readout PICNIC\cite{Kozlowski00, Cabelli00}) detector which receives near-infrared light (1.5 - 2.5 $\mu$m) from both interferometric outputs of the LBTI when the system is arranged for either nulling or Fizeau interferometry. PHASECam utilizes re-imaging optics to produce pupil images of each output beam, which are currently observed using standard $H$-band (1.65 $\mu$m) and $K$-band (2.2 $\mu$m) filters, respectively\cite{Defrere14}.

\begin{figure}[ht!]
\centering
\includegraphics[width=\textwidth]{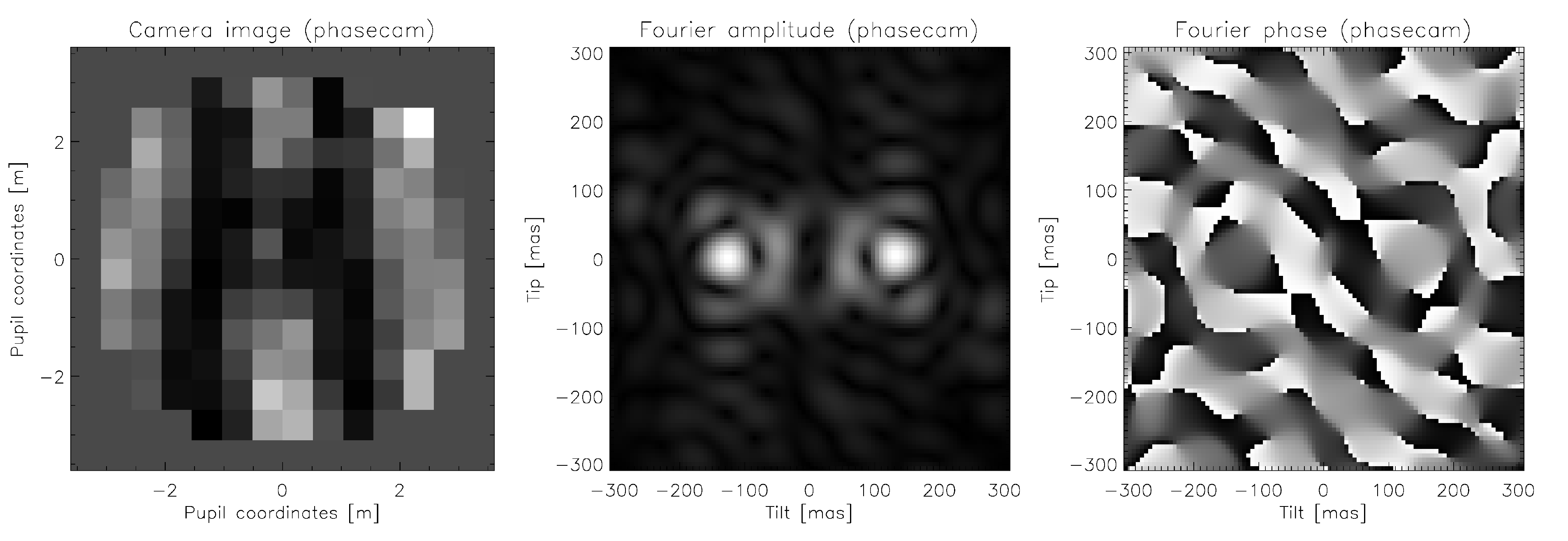}
\caption{LBTI's approach to phase sensing demonstrated for one interferometric output for on-sky $K$-band data taken on March 14, 2017. Pupil images of the interferometric output are imaged by PHASECam (left), and the Fourier transform of each image is used to perform tip/tilt and phase sensing. The position of the peak amplitude of the transform (middle) measures the differential tip/tilt. There are two peaks due to the dual-valued nature of the transform. The position is measured from one peak to the center of the pupil image: the argument of the transform at that position (right) measures the phase.\cite{Defrere14}}
\label{pupil-images}
\end{figure}

The current approach to phase sensing and control uses the $K$-band output pupil image. When the two input beams are well overlapped at the science wavelength of 10 $\mu$m for nulling, dispersion in the beamsplitter between 2 and 10 $\mu$m leads to a tilt difference of approximately three fringes across the pupil at 2 $\mu$m. This is intentional: it produces a signal in the Fourier plane which is well separated from the zero-frequency component. The differential tip/tilt and phase variations can be derived from a Fourier transform of this pupil image\cite{Defrere14}. This process is laid out in Fig. \ref{pupil-images}, which shows a $K$-band pupil image with fringes, and the amplitude and phase of the Fourier transform.

This measured ``raw" phase value is then processed to produce an OPD correction, which we describe further in the following section. During closed-loop operation the $K$-band phase measurements are translated to OPD corrections at a rate of 1 kHz and sent to the Fast and Slow Pathlength Correctors (FPC/SPC)  located in the UBC, which can also be seen in Fig. \ref{system-block}. Analogous calculations are also performed with the $H$-band output. The $H$-band measurements have been previously used to measure and correct for phase dispersion and water vapor variations between the two outputs\cite{Defrere16b}. However, this is currently unused as the impact of these effects turned out to be not as intractable as expected, which we discuss further in Sec. 4.

\subsection{A Closer Look at the Current Algorithm}\label{algorithm}

Here we describe the critical components of the phase measurement and OPD correction algorithm, with particular attention to the $K$-band implementation, so as to provide context for how fringe jumps occur and are currently corrected. 

\subsubsection{Raw Phase}

\begin{figure}[ht!]
\centering
\includegraphics[width=\linewidth]{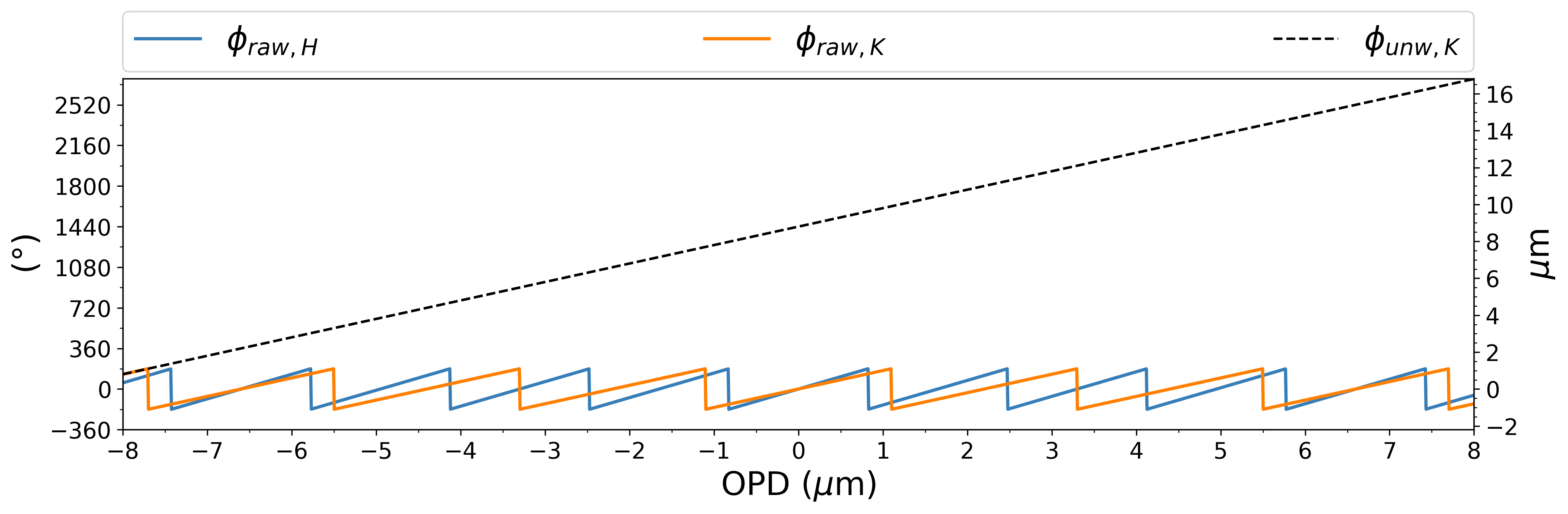}
\caption[angle=90]{The raw $H$-band and $K$-band phase values measured by PHASECam operating in open loop for an ideal (noiseless) linear OPD scan from -8 to 8 $\mu$m, and the associated unwrapped $K$-band phase in phase space (left axis) and delay space (right axis). If we assume 0 OPD is the fringe we want to lock onto, this shows how the unwrapped phase is a relative measurement: at 0 $\mu$m absolute OPD the unwrapped phase is not 0$\degree$ but instead $\sim$1400$\degree$, the distance away from the fringe in which the loop closed.} 
\label{raw-phases}
\end{figure}

The ``raw" $H$-band and $K$-band phases, $\phi_{H/K,raw}$, or the ``wrapped" phases, are the phase values measured from the Fourier transforms of the $H$-band and $K$-band pupil images, which can be seen in Fig. \ref{raw-phases}. They are restricted to a range of [-$180$, $180$)$\degree$ due to the cyclical nature of the transform. They indicate the position relative to the center of the current fringe, whichever fringe that may be.

\subsubsection{Unwrapped Phase}
\label{sect:unwph}

The unwrapped phase, $\phi_{K,unw}$ is the ``true" total differential phase between the two sides of the LBTI at the current time step, measured relative to the fringe on which the loop closed. To calculate this, we use a first order derivative algorithm\cite{Colavita10}. The basis of this algorithm is the assumption that OPD variations are of large amplitude but are slow --- they happen over a relatively long period of time as compared to the rate of correction. Thus, on small timescales, such as the 1 kHz correction rate of PHASECam, the derivative of the phase is of small magnitude. 

In practice, the unwrapped phase is initialized to the first raw phase value acquired after loop closure. To calculate the unwrapped phase value at each timestep $i$ thereafter, we use the relation
\begin{equation}\label{unwrapcheck}
\phi_{K,unw,i} - \phi_{K,unw,i-1} = (\phi_{K,raw,i} - \phi_{K,raw,i-1} + 180)\ \%\ 360 - 180.
\end{equation}
The delta in raw phase is the delta in unwrapped phase wrapped to between -180$\degree$ and 180$\degree$. The above equation adds or subtracts 360$\degree$ from the unwrapped phase value if the delta in the raw phase is bigger than 180$\degree$, assuming the actual delta is small and was wrapped. The unwrapped $K$-band phases for the previous linear pathlength scan can also be seen in Fig. \ref{raw-phases}.

\subsubsection{Correction and Residuals}
\label{corr}

The final value of the unwrapped phase is then used to calculate the applied OPD correction, using the expression,
\begin{equation}
    OPD = (s - \phi_{K,unw,i-1}) \times \frac{\lambda_{K}}{360},
\end{equation}

which is then sent to the FPC/SPC in the UBC. In this expression, $s$ is the pathlength setpoint, or the $K$-band position relative to the fringe on which the loop closed to which the pathlength correction loop is attempting to drive the system at every timestep. The setpoint is initially set to the position at which the loop is closed, and then typically iterated in small steps until the location of the deepest null (for nulling interferometry) or zero OPD (for Fizeau interferometry) is found.

\begin{figure}[t!]
\centering
\includegraphics[width=\textwidth]{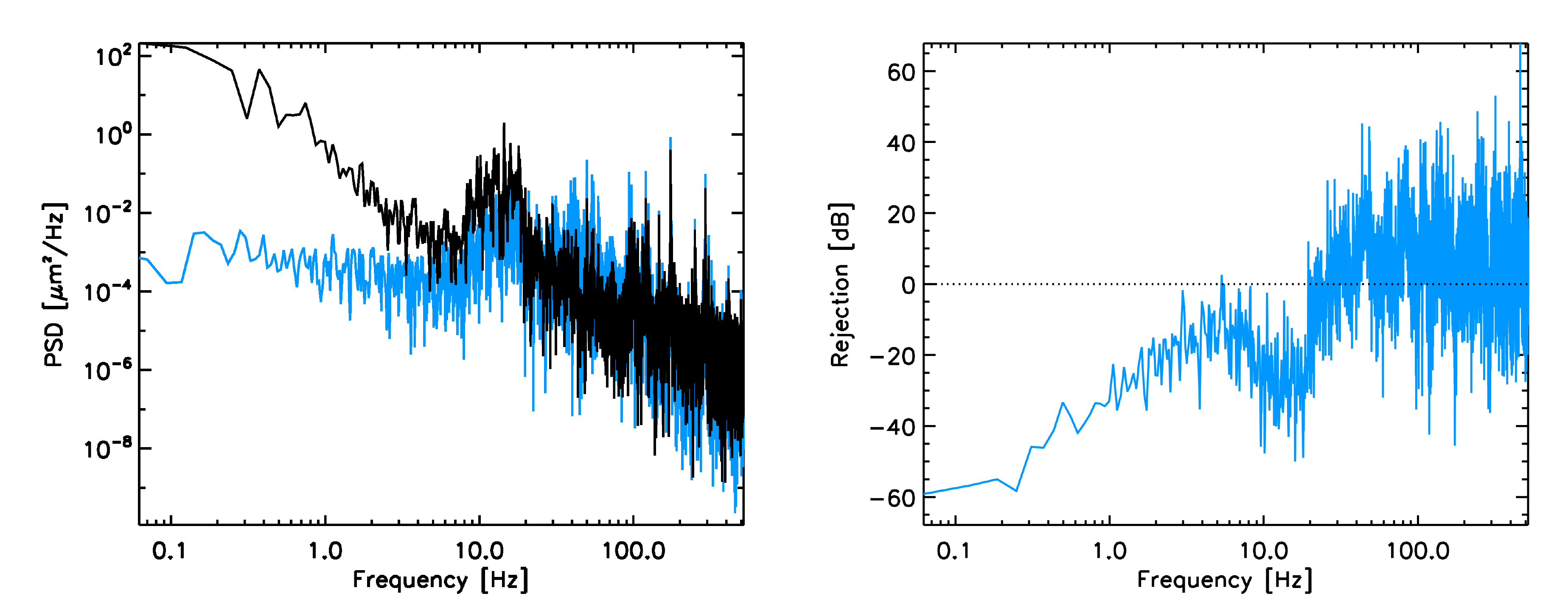}
\caption{$Left$: Power spectral densities of the differential OPD variations between the two AO-corrected LBT apertures in closed loop (blue) and open loop (black). Most OPD residuals come from high-frequency perturbations ($>$20 Hz). $Right$: Corresponding frequency response. Data obtained on February 4, 2015 on the bright star $\mu$ Gem. Figure adapted from Defr\`{e}re 2016a.\cite{Defrere16a}}
\label{example-psd}
\end{figure}

Even with the pathlength correction loop running at a rate of 1 kHz, PHASECam has a residual OPD of approximately 50-65$\degree$ (0.3 - 0.4 $\mu$m) rms\cite{Defrere16a}. While PHASECam is quite successful at removing the effects of instrument flexure ($\ll$1 Hz) and the atmosphere ($\sim$10 Hz), it is less successful at removing the effects of low frequency telescope resonances (12-18 Hz) and higher-frequency instrument vibrations (100-150 Hz)\cite{Defrere16a,Defrere16b} not already corrected for by the LBT's accelerometer network\cite{Kurster10, Bohm17}. The latter effect dominates the current residuals, which can be seen in Figure \ref{example-psd}.

\subsection{Fringe Jumps}

As peviously described, even at a 1 kHz correction rate, fringe jumps still occur. This may in some cases occur due to the atmosphere or vibrations causing an OPD variation larger than 180$\degree$, breaking the assumption of small phase variations previously described. However, given the relatively low magnitude of PHASECam's OPD residuals, a more likely culprit is an error in the phase estimation due to low signal-to-noise ratio (SNR). As shown in Figure \ref{pupil-images}, the precision of the phase estimation depends on the peak of the amplitude of the Fourier transform, which depends on the Strehl ratio delivered by the AO system. If the error in the phase estimation is large enough, this may lead to to a fringe jump or to the pathlength correction loop opening. More definitively quantifying the causes and rates of occurrence of fringe jumps requires a complex modeling effort outside the scope of this work.

\section{The Multi-wavelength Approach}

As previously described in Sec. \ref{overview}, PHASECam simultaneously measures phase telemetry in both the $H$ and $K$ bands. We can thus ``scaffold" the $K$-band telemetry with the $H$-band telemetry, removing the phase degeneracy out to the first common multiple of these wavelengths\cite{Meisner12}. This is 6.6 $\mu$m, or three fringes in $K$ band, which allows us to identify whether PHASECam is locked onto the correct fringe or one immediately adjacent to it.

\begin{figure}[ht!]
\centering
\includegraphics[width=\linewidth]{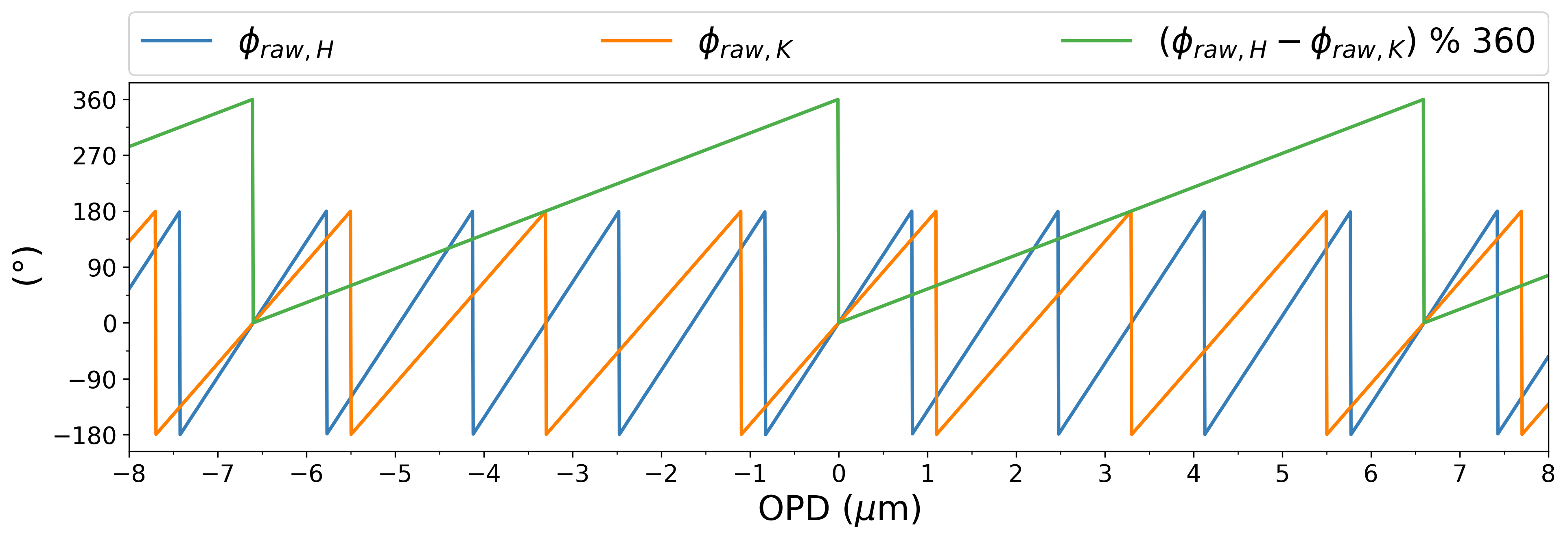}
\caption[angle=90]{The $H$- and $K$-band raw phases and the diffmod for the same ideal linear OPD scan. The diffmod wraps with a period of 3 $K$-band fringes.}
\label{raw-and-diff}
\end{figure}

\subsection{The Difference-Modulo Metric}

\noindent The core of this two-band approach to fringe jump capture and correction is conceptually based upon methods used by Meisner et al.\cite{Meisner12} for the Nova Fringe Tracker proposal for the VLTI. We call it the difference-modulo metric, henceforth referred to as the diffmod. The instantaneous diffmod per timestep, $d$, is generally mathematically described by
\begin{equation}\label{diffgen}
d = (\phi_{H,raw,i} - \phi_{K,raw,i})\ \%\ 360\degree,
\end{equation}
where \% represents the modulo, or remainder, operation.

The benefit of the modulo operation is illustrated in Fig. \ref{raw-and-diff} with the same linear phase scan as in Fig. \ref{raw-phases}. It produces a monotonic metric from 0$\degree$ to 360$\degree$ over the course of 6.6 $\mu$m OPD. This adequately captures the typical range of phase variations seen by PHASECam during closed-loop operation.

How does this metric allow us to detect fringe jumps? While $d$ may vary rapidly due to noise, so long as the system is still within the same PHASECam fringe, the time average of $d$, $<d>_{t}$, should be approximately constant. This is because PHASECam is constantly trying to drive the OPD to a single value; the pathlength setpoint $s$. It is therefore driving towards a single pair of raw phase values, and thus a single value of $<d>_{t}$. Under nominal circumstances --- i.e., no fringe jumps --- $<d>_{t}$ should only change if the setpoint does. To calculate $<d>_{t}$, we must use phasors, as $d$ is an angle which cannot be averaged directly. This is done using the formula 
\begin{equation}\label{timeaverage}
    <d>_{t} = arg(<e^{id}>_{t}) + 180\degree
\end{equation}
where the +180$\degree$ term shifts the range from the range of the arctangent function, [-180$\degree$,180$\degree$), back to the regular range of the diffmod.

\begin{figure}[t!]
    \centering
    \includegraphics[width=1.0\textwidth]{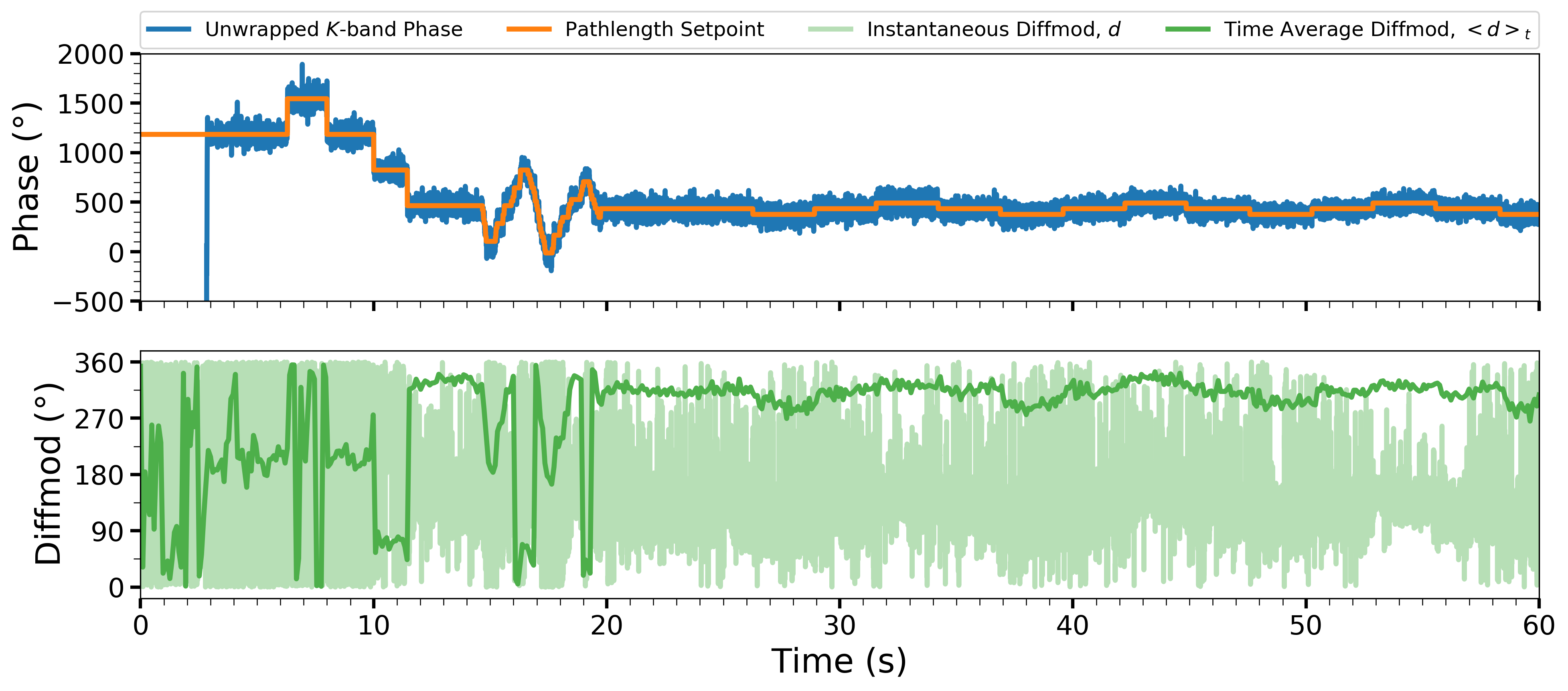}
    \caption{$Top$: The unwrapped $K$-band phase $\phi_{K,unw}$ and associated pathlength setpoint $s$ for a 1 minute PHASECam nulling telemetry sequence containing no fringe jumps. The pathlength correction loop closes at $\sim$3s. 360$\degree$ jumps in $\phi_{K,unw}$ and $s$ due to operator setpoint changes are visible at $\sim$6.5s, 8s, 10s, and 12s. These are followed by a sawtooth pattern that indicates that the null setpoint finding script was being run. The variations with a period of $\sim$10 s thereafter are due to automated dithering of $s$ in order to mitigate the effects of dispersion between K band and N band where nulling observations are performed. $Bottom$: The instantaneous diffmod $d$ overlaid by the average diffmod $<d>_{t}$. The averaging period is 100 phase values, or $\sim$0.1 s. It can be seen that $<d>_{t}$ generally well follows changes in $s$, with some anomalous spikes due to wrapping. The changes in $<d>_{t}$ due to operator setpoint changes are of approximately the magnitude predicted by Eq. \ref{predicteddiff}. The offset between $d$ and $<d>_{t}$ is due to the additive term in Eq. \ref{timeaverage}.}
     
    \label{rawdiffmodave}
\end{figure}

In Fig. \ref{rawdiffmodave} we show $d$ for an example PHASECam telemetry sequence containing no fringe jumps, overlaid by $<d>_{t}$, along with the associated $\phi_{K,unw}$ and $s$. It can be seen that $<d>_{t}$ generally stays close to the same value unless the pathlength setpoint is changed, as predicted. The noise in $d$ comes from a larger noise contribution from the $H$ band: the 50-60$\degree$ RMS residual OPD is a $K$-band measurement. The averaging period of $\sim$0.1 seconds is set to strike a balance between smoothing out this noise in $d$ and capturing the variations that are significant. With shorter averaging periods, many spurious spikes in the declared relative fringe value occur. 

We can predict the expected behavior of $<d>_{t}$ when a fringe jump occurs. If we express the $H$-band phase as a fraction of the $K$-band phase, the difference between the $H$ and $K$ phases is,
\begin{equation}
\phi_{H} - \phi_{K} = \left(\phi_{K} \times \frac{\lambda_{K}}{\lambda_{H}}\right) - \phi_{K} = \frac{4}{3}\phi_{K} - \phi_{K} = \frac{1}{3}\phi_{K}.
\end{equation}\label{predicteddiff}
By this relationship, a 360$\degree$ fringe jump in $K$ band, should lead to a fractional $\frac{1}{3}\Delta\phi_{K}$ change in $<d>_{t}$, or $\sim$120$\degree$. This should also occur with 360$\degree$ pathlength setpoint changes, which is confirmed by Fig. \ref{rawdiffmodave}. We thus now define the reference diffmod $R$. $R$ is the initial value of $<d>_{t}$ after the pathlength correction loop closes, to which each value of $<d>_{t}$ thereafter will be compared to determine whether or not a fringe jump has occurred.

\subsection{The Diffmod Loop}

We now describe the conceptual implementation of the diffmod in the PHASECam system. This involves five main steps, which can be seen graphically in Fig. \ref{diffmodloop}. They are:

\begin{figure}[ht!]
\centering
\includegraphics[width=\textwidth]{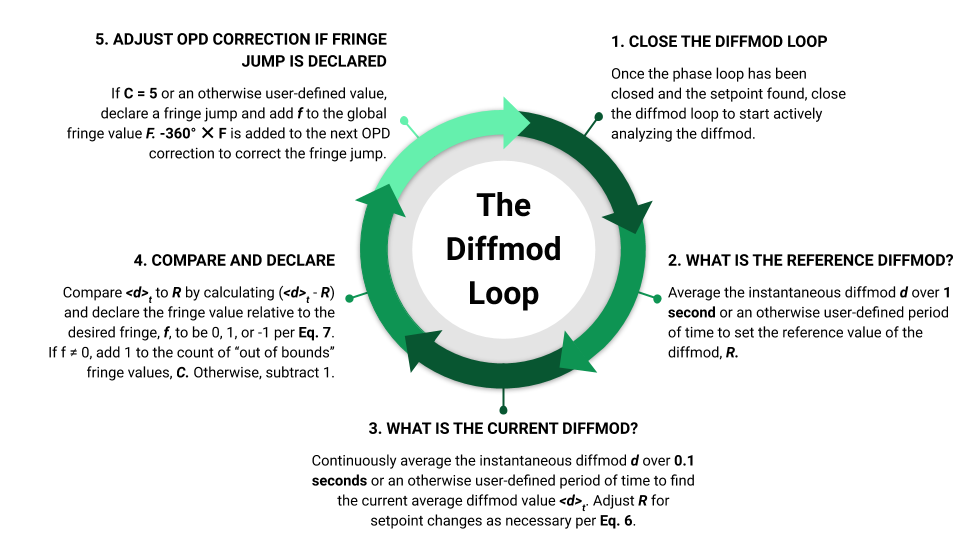}
\caption[angle=90]{The five core steps of the diffmod algorithm.}
\label{diffmodloop}
\end{figure}

\subsubsection{Close the diffmod loop} 

Similarly to the pathlength correction loop, we conceive the diffmod algorithm as a loop structure with ``open" and ``closed" states. In the open state $d$ is still calculated at the full 1 kHz rate, but only in the closed state is $<d>_{t}$ calculated and fringe jumps tracked/declared. These states for the diffmod loop are not tied to the state of the main pathlength correction loop. For maximum accuracy it is best to wait to close the diffmod loop until after we have found the fringe we intend to lock onto. 

\subsubsection{Calculate the diffmod reference value $R$}

 After the diffmod loop closes, we average $d$ over 1 second to produce $R$. If $s$ changes at any point where a fringe jump has not occurred after $R$ has been calculated, then the value of $R$ should be adjusted by the relation
\begin{equation}
R' = \left(R + \left[(s' - s) \times \frac{\lambda_{K} - \lambda_{H}}{\lambda_{K}}\right]\right) \%\ 360\degree,
\end{equation}\label{new-ref-eq}
where $R'$/$s'$ are the new values of the reference diffmod and the pathlength setpoint. The final multiplicative factor in the expression is $\frac{1}{3}$, due to the previously described scaling relation between the $K$-band and $H$-band wavelengths. The modulo accounts for the possible wrapping of $R$.

\subsubsection{Calculate the current average diffmod $<d>_{t}$}

Once $R$ has been calculated, we begin calculating $<d>_{t}$ over a shorter user-defined period of time. We again use 100 $d$ values --- approximately 0.1s --- as our averaging period, which keeps the time to declared detection of a fringe jump relatively small while again smoothing out spurious variations. Depending on observing conditions, this averaging period could possibly be reduced or may need to be extended.

\subsubsection{Compare $<d>_{t}$ to $R$ and declare the current fringe}

We now calculate the difference between $<d>_{t}$ and $R$, and declare whether the system is still in the desired fringe or $\pm$1 fringe per Eq. \ref{fringeid}. We set $\pm$60$\degree$ as the threshold for exiting the desired fringe. Each fringe is 120$\degree$ wide in diffmd space, thus 60$\degree$ is the minimum boundary for transition into another fringe. Thus, to declare the relative fringe value, $f$, we use the scheme:

\begin{equation}\label{fringeid}
f\ = \left\{
\begin{array}{ll}
      -1 & <d>_{t} - R\ \leq\ -60\degree\ or \geq 180\degree\\ 
      0 & |<d>_{t} - R|\ <\ 60\degree \\
      1 & <d>_{t} - R\ \geq\ 60\degree\ or \leq -180\degree\\
\end{array} 
\right. 
\end{equation}

The conditional statements represent the scenario where $<d>_{t}$ may cross the wrap and end up $\sim$240$\degree$ away from $R$ rather than $\sim$120$\degree$, where 180$\degree$ is the minimum boundary for transition. We make the assumption that a singular fringe jump in one direction is much more likely than a double fringe jump in the other direction.

We require the diffmod loop to declare a non-zero fringe a minimum number of averaging periods in a row before positively declaring a fringe jump and activating the correction step,  in order to be robust against short drifts into or out of a fringe and random noise spikes. The count of ``out-of-bounds" (non-zero) fringe values, $C$, is kept on a sliding scale for similar reasons. An out-of-bounds value adds 1 to $C$, and an in-bounds value (zero fringe) subtracts 1. Currently, the minimum is 5.

\subsubsection{Correct the fringe jump}

\noindent If correction of a fringe jump is required, 360$\degree$ will be added to or subtracted from the next applied OPD correction. Once the system has returned to the correct fringe, the out-of-bounds count will begin counting down to zero again, at which point the current fringe jump will be declared corrected. The diffmod loop will then return to the monitoring state. If there has been more than one fringe jump, this process repeats for each individual fringe jump, until the global relative fringe value returns to zero.

\subsection{Testing the Diffmod Algorithm}

We have tested the diffmod algorithm utilizing archival PHASECam telemetry. We constructed an analysis pipeline which scans through telemetry, searching for fringe jumps using the diffmod framework. Each time the pathlength correction loop closes for more than 5 seconds, the user is asked to identify when they would like the diffmod loop to close relative to that point. Analysis then proceeds as described in the previous section.

We can tell when a fringe jump may have occurred by looking at 360$\degree$ changes in the pathlength setpoint in the unwrapped phase telemetry. A mono-directional setpoint change is the hallmark of a fringe jump correction. That is, a fringe jump most likely occurred shortly beforehand if a) the pathlength setpoint changes by 360$\degree$ in one direction and does not change back within a few seconds, or b) changes by 360$\degree$ in one direction, changes back and then changes by 360$\degree$ in the other direction and remains there. These setpoint changes are applied by the PHASECam operator to send the system back to the original fringe: in the second case, the operator corrected the setpoint in the wrong direction first. A setpoint change that quickly returns to the original value is not a fringe jump. The setpoint is sometimes briefly dithered by the operator if the loop appears to be becoming unstable, distinct from the automated dithering performed during nulling observations.

We have analyzed a continuous 8 minute telemetry sequence. The telemetry is from observations of HD168775, a HOSTS calibrator star, ($K$ $\sim$1.74) taken on March 28, 2018 during the last of the primary HOSTS observations. Live on-sky testing has yet to be performed primarily due to limited availability of closed pathlength loop observations (there has been a low number of these programs in the LBTI queue due to the completion of the primary HOSTS observations and the commissioning of controlled Fizeau interferometry still being in progress).

\section{Results and Discussion}\label{results}

\subsection{Results of Archival Telemetry Analysis}

\begin{figure}[H]
    \adjustimage{width=.95\textwidth,left}{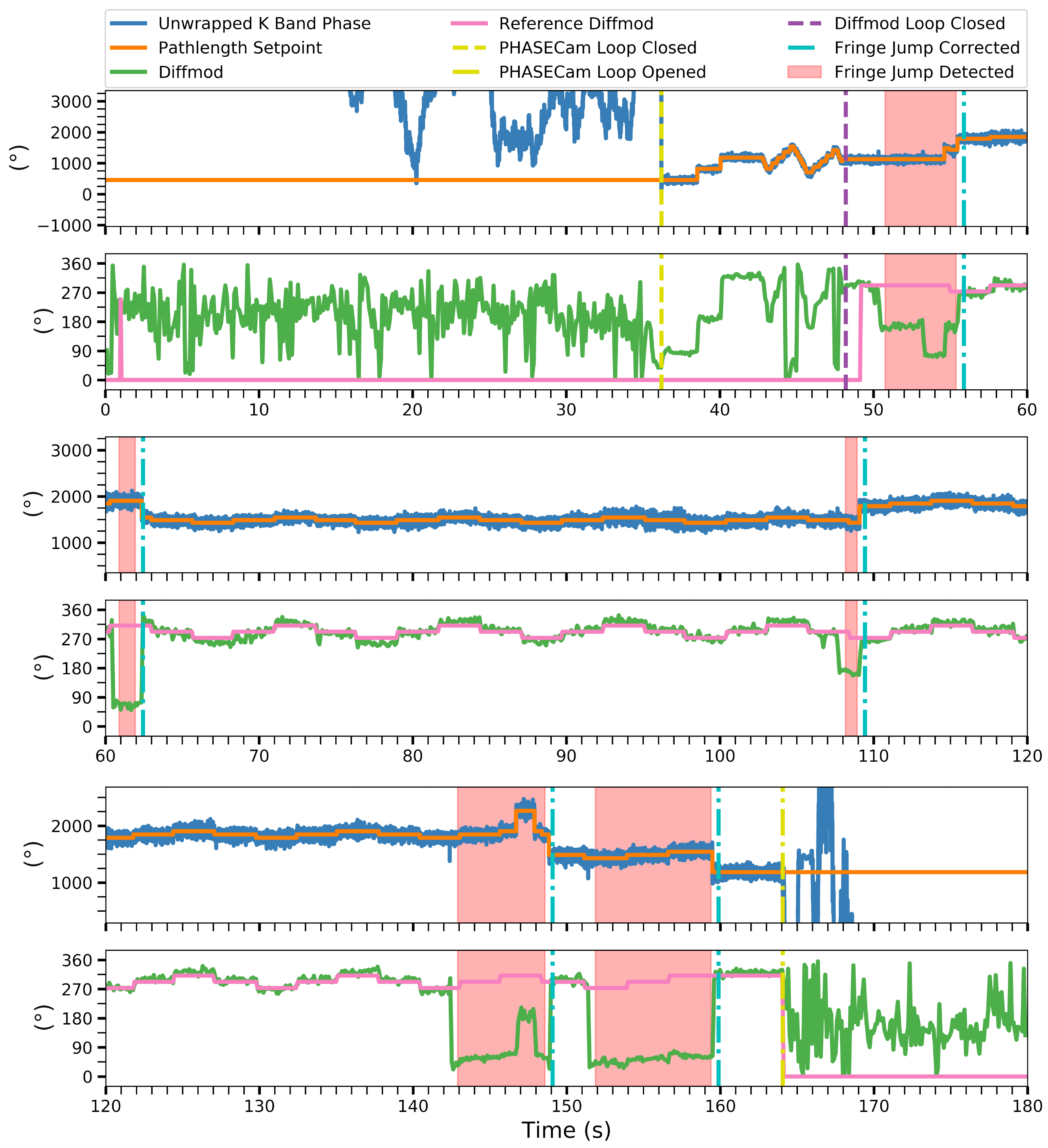}
    \vspace{.1mm}
    \caption{Diffmod analysis of UT 03/28/18 10:40-10:42. The pathlength correction loop closes at 0:36. The diffmod loop closes at 0:48 after the setpoint saw-tooth (0:43 - 0:48). Two +360$\degree$ setpoint changes at 0:54.5 and 0:55.5 indicate two negative fringe jumps, confirmed in the diffmod analysis by successive decreases in $<d>_{t}$ at $\sim$0:51 (-1, 1) and 0:53.5 (-2, 2). Similarly, fringe jumps occur at 0:60.5 (+1, 3), 0:107.5 (-1, 4), 0:142.5 (+1, 5) and 0:151.5 (+1, 6). Fringe jumps (3), (5), and (6) are +1 fringe jumps per Eq. \ref{fringeid}: the $\sim$240$\degree$ change in $<d>_{t}$ indicates a +1 fringe jump that crossed the wrap. The first setpoint change sent to correct (5) is in the wrong direction: $<d>_{t}$ does not return to $R$. A setpoint change in the reverse direction at 0:149 properly corrects it. All of fringe jumps (1-6) were declared detected and corrected after $\sim$0.5s, as designed. At $\sim$0:164, the pathlength correction loop opens.}
    \label{diffsequences1}
\end{figure}

Figure \ref{diffsequences1} demonstrates a representative 3 minute sub-sample of the results of our archival telemetry analysis (UT 03/28/2018 10:40-10:42), from first pathlength correction loop closure to opening. $\phi_{K,unw}$, $s$, $R$ and $<d>_{t}$ are displayed. For demonstration purposes, $<d>_{t}$ is continuously calculated whether or not the diffmod loop or pathlength correction loop has been closed. Pertinent telemetry features are detailed in the caption. Fringe jumps are introduced with their timestamp and a parenthetical containing their relative fringe value and ID number per Table \ref{jumptable}. Table \ref{jumptable} lists the ID number of all features we identified fringe jumps in the 8 minute sequence, their timestamp, the relative fringe value, and the time lost to the jump or cluster of jumps to the nearest half second.

\begin{table}[t]
\begin{center}
\begin{tabular}{|c|c|c|c|}
\hline
\textbf{Jump ID} & \textbf{Time of Occurrence (s)} & \textbf{Fringe Value} & \textbf{Time Lost (s)} \\ \hline
1                & 51                              & -1                    & \multirow{2}{*}{5}   \\ \cline{1-3}
2                & 53.5                            & -2                    &                        \\ \hline
3                & 60.5                       & +1                    & 2                    \\ \hline
4                & 107.5                           & -1                    & 2                    \\ \hline
5                & 142.5                           & +1                    & 7                    \\ \hline
6                & 151.5                           & +1                    & 8.5                      \\ \hline

7                & 245.5                           & +1                    & \multirow{3}{*}{4.5}   \\ \cline{1-3}
8               & 247.5                     & +2 & \\ \cline{1-3}

9               & 248.5                     & +1 & \\ \hline

10                & 257                           & +1                    & \multirow{4}{*}{6}   \\ \cline{1-3}

11                & 258.5                          & +2                    &   \\ \cline{1-3}

12                & 259.5                           & +1                    &   \\ \cline{1-3}
13                & 262                           & +1                   &                        \\ \hline
14\$               & 282                             & -1                    & \multirow{2}{*}{5} \\ \cline{1-3}     
15\$              & 286.5                 & 0 &  \\ \hline
16               & 292.5                           & -1                    & \multirow{2}{*}{4.5}     \\ \cline{1-3}
17               & 293.5                           & -2                   &                        \\ \hline

18               & 298                           & -1                   & 2                       \\ \hline
19               & 309                             & +1                    & \multirow{2}{*}{34.5}    \\ \cline{1-3}
20\#               & 310                             & +2                    &                        \\ \hline
21               & 366                             & +1                    & \multirow{3}{*}{21}  \\ \cline{1-3}
22               & 368                             & +2                    &                        \\ \cline{1-3}
23\#               & 369                             & +3                   &                        \\ \hline
24\$               & 397                             & +1                    & \multirow{2}{*}{4.5}     \\ \cline{1-3}

25\$              & 401                   & 0 & \\ \hline

26               & 411                             & +1                    & \multirow{3}{*}{11.5}    \\ \cline{1-3}
27               & 416                             & +2                    &                        \\ \cline{1-3}
28               & 417                             & +3                   &                        \\ \hline
29\#               & 431                             & +1                    & 29.5                     \\ \hline
\end{tabular}
\end{center}
\caption{List of fringe jump events in the full 8 minute telemetry sequence. \$ - pairs of fringe jumps which cancelled each other out and were not detected by the operator; \#: fringe jumps that were not corrected before the pathlength correction loop opened and thus the time before next loop closure is included in time lost.}
\label{jumptable}
\end{table}

\subsection{Discussion}\label{discuss}

\subsubsection{Completeness and Accuracy of Fringe Jump Detection}

The diffmod algorithm is currently designed for detection of fringe jumps only of magnitude 1. This is an intentional decision made under the assumption that under median conditions fringe jumps occur at a rate that will allow ``clusters" like those in Table \ref{jumptable} to be broken up into individual detection and correction events.

Of the 15 ``primary" fringe jumps, i.e., those individual jumps of magnitude 1 or the first jump in a cluster the diffmod had a 100\% rate of detection and declaration of relative fringe value. This includes 2 fringe jumps (14 and 24) which were not detected by the operator and ``self-corrected" via a fringe jump in the opposite direction.

\begin{figure}[H]
    \adjustimage{width=\textwidth}{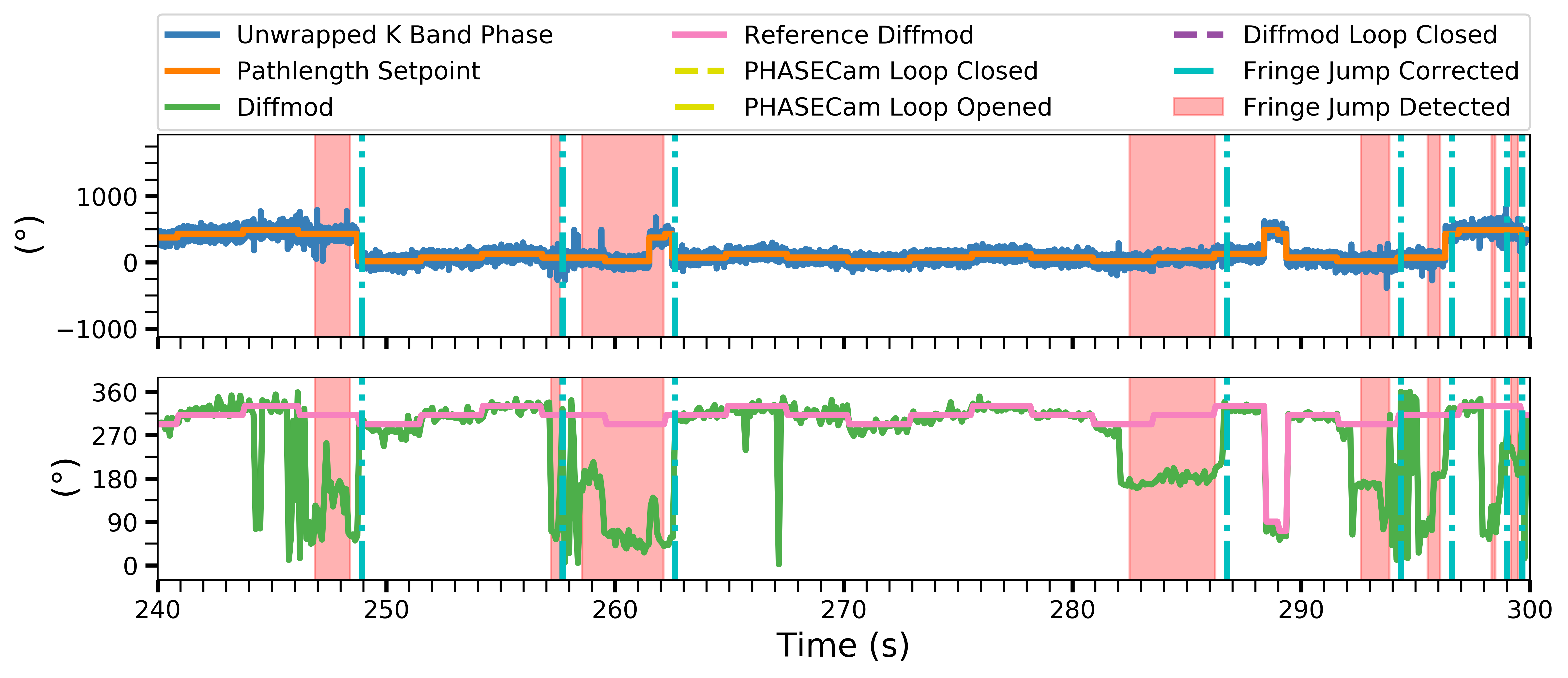}
    \vspace{.1mm}
    \caption{Diffmod analysis of UT 03/28/18 10:44. This sequence contains jumps (7) - (18), per Table \ref{jumptable}. $<d>_{t}$ is very noisy in several places, causing delays in fringe jump detection or in some cases spurious declarations of correction. Attempts by the operator to stabilize the loop by briefly dithering the setpoint can be seen at 0:261.5 and 0:288.5. See the text for further discussion.}
    \label{noisy-telemetry}
\end{figure}

Still, on-sky testing will be required to determine how well the above assumption performs. Several of the jump clusters have very close occurrence times between constituent fringe jumps. These typically occurred under particularly unstable conditions in this sequence. This can be seen in Figure \ref{noisy-telemetry}, a later segment of the 8 minute sequence. In some cases in this sequence, the diffmod took a longer time to declare a fringe jump or prematurely declared a correction due to noise in $<d>_{t}$. 

\subsubsection{Quantification of Time Lost}

Time from spike in $<d>_{t}$ to declared detection and time from return to $R$ to confirmation of correction both typically occurred within 0.5s, as designed. Correction in the on-sky implementation of the diffmod algorithm is expected to take negligible time, as it is an adjustment to a single OPD correction. Thus, we estimate a full duty cycle from initial occurrence of a fringe jump to confirmed correction of $\sim$1 s. Based on this sequence however, particularly Fig. \ref{noisy-telemetry}, this may sometimes be extended depending on observing conditions. However, this is a significantly shorter time interval than manual correction achieves on average, as seen in this sequence where manual correction regularly took $\sim$ 5s or longer.

We quantify the impact of fringe jumps on this data set by defining two metrics - ``actual"  time lost vs. ``potential" time lost. Actual time lost refers to the time lost to fringe jumps while the pathlength correction loop is closed. ``Potential" time lost refers to instances where fringe jumps occurred and were not corrected before the pathlength correction loop opened. In these cases we count all time between the pathlength correction loop opening and re-closing as ``potential" science time lost to fringe jumps. Any time lost to these fringe jumps prior to the loop opening is then subsumed into the ``actual" time lost tally. 

By these metrics, $\sim$139 seconds of time were lost to fringe jumps, or approximately 29\% of the 480s sequence. We expect that a fair fraction of this lost time could be recovered with implementation of the diffmod algorithm. A non-negligible fraction of phase loop breakages are caused by slowly and/or wrongly corrected fringe jumps. The diffmod may be able to significantly reduce the rate of loop breakage due to this, as well as reduce the amount of time off-fringe and thus data lost, by turning multiple-jump events into discrete single jump events.

\subsubsection{Water Vapor Dispersion}

In addition to the abrupt phase variations caused by fringe jump, another source of phase changes that PHASECam sees is water vapor seeing\cite{Colavita04}. Variations in the differential water vapor in the atmosphere above the telescope apertures can cause slow variations in the fringe chromatic phase, and thus in the diffmod, on timescales of several seconds. While it is unlikely that water vapor can be implicated as the sole cause of any singular jump event, it certainly contributes to the overall behavior of the diffmod, and may be partially responsible for the residual noise in $<d>_{t}$ which causes effects like those seen in Fig. \ref{noisy-telemetry}.

In general, the diffmod appears to be able to serve as a qualitative check on the stability of the loop and observing conditions. However, closing the diffmod loop is likely to produce diminishing returns on overall loop stability under poor conditions, as time to fringe jump detection will likely increase beyond the typical time between successive fringe jumps.

\section{Summary and Conclusions}

We have introduced the difference-modulo metric, or diffmod, a method based on concepts utilized by Meisner et al. (2012)\cite{Meisner12} for the Nova Fringe Tracker concept for the VLTI. The diffmod simultaneously uses the $H$ and $K$-band raw phase telemetry available to us from LBTI's fringe tracker, PHASECam, thus breaking the degeneracy of the phase delay over a range of 6.6 $\mu$m. In this way the telemetry can be used to automatically detect and correct fringe jumps which degrade the stability of the pathlength correction loop and reduce achievable visibilities. The diffmod is independent of observing mode as it only utilizes the raw phase measurements.

Through analysis of archival telemetry, we have shown the diffmod is a simple yet viable method of fringe jump detection. It successfully detected all ``primary" fringe jumps in our analyzed sequence on average much faster than the operator. It thus has the potential to recover a significant fraction of the science time lost to fringe jumps. Remaining frames which are in the incorrect fringe can be filtered out in post-processing through the addition of a few diffmod-related variables to the recorded telemetry. 

In addition, the diffmod is useful as a secondary qualitative check on the stability of the pathlength correction loop and observing conditions. This includes the water vapor dispersion, which can present as slow variations in the diffmod and contribute to residual noise. Having a ``live" plot of the diffmod available to PHASECam operators may thus prove useful. The diffmod algorithm will be implemented as a closeable loop independent of the pathlength correction loop such that the operator may revert to manual correction if necessary.

The next steps for the diffmod algorithm are integration into the PHASECam codebase and robust on-sky testing at the full 1 kHz data rate with active correction of fringe jumps. The  limits of the diffmod's detection thresholds as a function of observing conditions must be explored, particularly its ability to dis-aggregate ``clusters" of fringe jumps as conditions deteriorate.

\acknowledgments 
LBTI is funded by a NASA grant in support of the Exoplanet Exploration Program (NSF 0705296). The LBT is an international collaboration among institutions in the United States, Italy and Germany. LBT Corporation partners are: The University of Arizona on behalf of the Arizona university system; Istituto Nazionale di Astrofisica, Italy; LBT Beteiligungsgesellschaft, Germany, representing the Max-Planck Society, the Astrophysical Institute Potsdam, and Heidelberg University; The Ohio State University, and The Research Corporation, on behalf of The University of Notre Dame, University of Minnesota and University of Virginia. The authors would like to thank the reviewers for their time and valuable feedback, particularly their feedback regarding the averaging of phasors. Thanks also to Jordan Stone for his insights regarding non-redundant aperture masking. This manuscript is an updated version of work presented by Erin Maier at the 2018 SPIE Astronomical Telescopes and Instrumentation conference in Austin, Texas\cite{Maier18}. Portions of this work were supported by the Arizona Board of Regents Technology Research Initiative Fund (TRIF). This work also made use of various Python packages for computing and plotting, including Matplotlib \cite{Hunter07}, NumPy \cite{Oliphant06}, IPython \cite{Perez07}, Jupyter notebooks \cite{Kluyver16a}, pandas \cite{McKinney10,pandas2020}, and plotly \cite{plotly}. It also made use of INDI, the Instrument-Neutral Distributed Interface, an open source architecture for control and automation of astronomical devices \cite{Downey07}.


\bibliography{report}   
\bibliographystyle{spiejour}   


\subsection*{Biographies}

\vspace{2ex}\noindent\textbf{Erin R. Maier} (they/them) is a doctoral candidate in astronomy at the University of Arizona. Their current research focuses on small development of instrumentation for high-contrast imaging from space and optical/infrared studies of debris disks. They received their master's degree in astronomy from the University of Arizona in 2019 and their bachelor's degree in astronomy and physics from the University of Iowa in 2017.

\vspace{2ex}\noindent\textbf{Denis Defr\`{e}re} obtained his PhD in Astrophysics from the University of Li\`{e}ge (Belgium) in 2009. After a post-doctoral stay at the Max Planck Institute for RadioAstronomie in Bonn, he was instrument scientist of the LBTI between 2012 and 2016. He was responsible for the nulling data pipeline and contributed to the commissioning of PHASECam. He is now a research associate at the University of Li\`{e}ge.

\vspace{2ex}\noindent\textbf{Steve Ertel} is an adaptive optics scientist at the Large Binocular Telescope (LBT) Observatory and leading instrument scientist for the LBT Interferometer (LBTI). His main expertise is in high contrast, high angular resolution observations including precision optical long baseline interferometry at infrared wavelengths, as well as in instrument science operations. He received his PhD from the University of Kiel, Germany, in 2012 and has held positions at the University of Grenoble, France, and at ESO Chile.

\vspace{2ex}\noindent\textbf{Katie Morzinski} is an Assistant Astronomer at Steward Observatory, Department of Astronomy, University of Arizona. Her research specializes in Adaptive Optics (AO) instrumentation for characterizing exoplanets through direct imaging, including contributions to AO systems and instruments at Lick Observatory, Gemini South, the Large Binocular Telescope, and the Magellan II telescope at Las Campanas Observatory. She has contributed to the LBTI instrument and observing teams since 2013.

\vspace{2ex}\noindent\textbf{Ewan S. Douglas} is an Assistant Professor of Astronomy at the University of Arizona Steward Observatory. His research focuses on space instrumentation, wavefront sensing and control, and high-contrast imaging of extrasolar planets and debris disks. He received his master's and doctoral degrees in astronomy from Boston University and his bachelor's degree in physics from Tufts University.

\vspace{2ex}\noindent Biographies of the other authors are not available.

\end{spacing}
\end{document}